\documentclass [superscriptaddress,amsmath,prl,twocolumn,amssymb]{revtex4-1}
\usepackage{graphicx}
\usepackage{dcolumn}
\usepackage[none]{hyphenat} 
\usepackage{braket}
\usepackage{multirow}
\usepackage{booktabs}
\usepackage{bm,color,upgreek}
\usepackage{braket}

\usepackage{graphicx}
\usepackage{dcolumn}
\usepackage{bm}

\begin{document}

\title{Tight-binding piezoelectric theory and electromechanical coupling correlations for transition metal dichalcogenide monolayers}

\author{Yunhua Wang}
\email{wangyh49@mail.sysu.edu.cn}
\affiliation{Sino-French Institute of Nuclear Engineering and Technology, Sun Yat-sen University, Zhuhai 519082, China}

\author{Zongtan Wang}
\thanks{Z. W. contributed equally to this work as Y. H. W..}
\affiliation{School of Engineering, Sun Yat-sen University, Guangzhou 510006, China}

\author{Jie Li}
\affiliation{School of Engineering, Sun Yat-sen University, Guangzhou 510006, China}

\author{Jie Tan}
\affiliation{Sino-French Institute of Nuclear Engineering and Technology, Sun Yat-sen University, Zhuhai 519082, China}

\author{Biao Wang}
\email{wangbiao@mail.sysu.edu.cn}
\affiliation{Sino-French Institute of Nuclear Engineering and Technology, Sun Yat-sen University, Zhuhai 519082, China}
\affiliation{State Key Laboratory of Optoelectronic Materials and Technologies, School of Physics, Sun Yat-sen University, Guangzhou 510275, China}

\author{Yulan Liu}
\email{stslyl@mail.sysu.edu.cn}
\affiliation{School of Engineering, Sun Yat-sen University, Guangzhou 510006, China}



\begin{abstract}
The lack of inversion symmetry in semiconducting transition metal dichalcogenide monolayers (TMDMs) enables a considerable molecular-level intrinsic piezoelectricity, which opens prospects for atomically-thin piezotronics and optoelectronics. Here, based on the tight-binding (TB) approach and Berry phase polarization theory, we establish an atomic-scale TB theory for demonstrating piezoelectric physics in TMDMs. Using the TB piezoelectric theory, we predict their electronic Gr\"{u}neisen parameters (EGP) which measure the electron-phonon couplings. By virtue of the constructed analytical piezoelectric model, we further reveal the correlation between the electronic contribution to piezoelectric coefficients and strain-induced pseudomagnetic gauge field (PMF). These predicted EGP and PMF for TMDMs are experimentally testable, and hence the TB piezoelectric model is an alternative theoretical framework for calculating electron-phonon interactions and PMF.
\end{abstract}

\pacs{Valid PACS appear here}
\maketitle

\textit{Introduction.} Since its first observation in 1880, piezoelectricity has been one of the most active topics in physics, because of its fascinating fundamental theory and wide applications in diverse fields. The striking features of piezoelectricity include the linear electromechanical coupling, reversibility and robustness against perturbations. It is a fact that the piezoelectric polarization difference contains both electronic and ionic contributions, because of the ionic internal-strain induced by the macroscopic deformation as well. In a microscopic picture, the piezoelectricity contributed by ions is associated with the Born effective charges and the optoacoustic coupling in the context of lattice-dynamical theory; the electronic polarization difference in response to the strain (or stress) field and the electronic part of piezoelectric coefficient are correspondingly related to the Berry phase \cite{polarization1,polarization2,polarization3} and the first Chern form \cite{chernDroth,chernLiWang,chernRostami}, which clarify the reason why is robust like the topological quantum states \cite{topologicalTKNN,Topologicalinsu}. Recently, the density functional theory (DFT) calculations \cite{piezoFP2dduerloo,piezoFP2dBlonsky,piezoFPTMDC1,piezoFPTMDC2,piezoFPgroupIVMD, piezoFPgraphene}, lattice dynamics calculations \cite{piezolattdynamical} and experiments \cite{piezoExpTMDCZLWang,piezoExpTMDCXZhang,piezoExpTMDCtotal, piezoExpGraphene,piezoExpC3N4,piezoExpIn2Se3} on the piezoelectricity of two-dimensional materials, particularly transition metal dichalcogenide monolayers (TMDMs) \cite{piezoFP2dduerloo,piezoFP2dBlonsky,piezoFPTMDC1,piezoFPTMDC2, piezolattdynamical,piezoExpTMDCZLWang,piezoExpTMDCXZhang,piezoExpTMDCtotal}, group IV monochalcogenides \cite{piezoFPgroupIVMD}, graphene \cite{piezoFPgraphene,piezoExpGraphene}, C$_3$N$_4$ \cite{piezoExpC3N4} and $\alpha$-In$_2$Se$_3$ \cite{piezoExpIn2Se3}, have shed new light on atomically-thin piezotronics, flexible electronics and optoelectronics.

TMDMs are semiconductors with experimentally tunable carrier mobilities \cite{TMDmobilities} and a direct band gap \cite{TMDgap}, which allows the field-effect transistors with a high on/off ratio \cite{TMDtransistor}. The absence of inversion symmetry results in a spin-orbit coupling (SOC) which lifts the spin degeneracy \cite{TMDSOC}, and the time-reversal symmetry keeps the valley degeneracy but makes the spin-splitting at different valleys opposite \cite{TMDValleyop}. The exotic spin-valley coupling together with the strong excitonic effect offers an avenue toward the valleytronics and optoelectronics \cite{TMDValleytronics}. The unique crystal structure and time-reversal symmetry are also responsible for the electromechanical couplings in TMDMs: (i) the trigonal prismatic structure with the main $d$ orbital interactions of the transition metal atoms leads to the high stiffness and breaking strength \cite{TMDstiff}, benefiting the nanomechanical resonators \cite{TMDnanomech} and flexible devices \cite{flexelectronic}; (ii) the broken inversion symmetry renders the piezoelectricity \cite{piezoExpTMDCZLWang,piezoExpTMDCXZhang}; (iii) the time-reversal symmetry enables the strain-induced valley-contrasting pseudomagnetic vector potentials \cite{ValleyPseuMagnetic1,ValleyPseuMagnetic2,ValleyPseuMagnetic3,ValleyPseuMagnetic4}, which measure the valley displacement analogous to that in strained graphene \cite{ValleyPseuMagneticGphe}; (iv) the lattice deformation modifies the electron-phonon interactions, which affect the electronic transport \cite{epEtransport1,epEtransport2,epExp}, optical properties \cite{epOptical}, spin relaxation \cite{epSpin} and valley magnetization \cite{epValleyMagnet}. Although much progress on the piezoelectricity and strain effects on the phonon \cite{strainEffphonon1,strainEffphonon2}, electronic properties \cite{strainElectron}, optoelectronic properties \cite{strainOptEle}, and work function \cite{strainWorkFuc} have been recently made by DFT and experiments, a microscopic theory, presenting the piezoelectric physics and revealing the correlation among these electromechanical couplings in TMDMs, is still lacking to date. In this Rapid Communication, we establish a microscopic piezoelectric theory using a combination of the tight-binding (TB) approach and Berry phase polarization theory, namely, TB piezoelectric theory. Using the linear feature of piezoelectricity and the correspondence between the TB piezoelectric model and the clamped-ion DFT model, we obtain the electronic Gr\"{u}neisen parameter (EGP), which characterizes the electron-phonon coupling in TMDMs. By virtue of the analytical piezoelectric model, we further explore the correlation between the electronic part of piezoelectric coefficient and strain-induced pseudomagnetic gauge field (PMF) and finally predict the PMF's values for strained TMDMs.

\textit{TB Hamiltonian and electron-phonon interactions.} For TMDMs (MX$_2$, M = Mo, W; X = S, Se, Te), the conduction and valence bands near the Fermi energy are mainly contributed by $d_{z^2}$, $d_{xy}$ and $d_{x^2-y^2}$ orbitals of M atoms \cite{dorbitals1,dorbitals2}. Consequently, their low-energy physics can be captured by the TB Hamiltonian including the nearest-neighbor (NN), next-nearest-neighbor (NNN) and third-nearest-neighbor (TNN) $d-d$ hoppings. In the real space, the TB Hamiltonian for unstrained TMDMs reads
\begin{eqnarray}
 H_0 = \sum_{i,\zeta}\epsilon_\zeta c^\dag_{i,\zeta}c_{i,\zeta}+\sum_{i,\bm\delta}\sum_{\zeta,\zeta^\prime} t_{\zeta,\zeta^\prime}c^\dag_{i,\zeta}c_{i+\bm\delta,\zeta^\prime}\nonumber \\
 +\sum_{i,\bm\chi}\sum_{\zeta,\zeta^\prime} r_{\zeta,\zeta^\prime}c^\dag_{i,\zeta}c_{i+\bm\chi,\zeta^\prime} +\sum_{i,2\bm\delta}\sum_{\zeta,\zeta^\prime} u_{\zeta,\zeta^\prime}c^\dag_{i,\zeta}c_{i+2\bm\delta,\zeta^\prime},
\end{eqnarray}
where $\epsilon_\zeta$ is the on-site energy, $c^\dag_{i,\zeta}$ and $c_{i,\zeta}$ are the creation/annihilation operators for an electron with the orbital $\zeta$ ($d_{z^2}$, $d_{xy}$ or $d_{x^2-y^2}$ ) on site $\bm{R}_i$, $\bm\delta$, $\bm\chi$ and $2\bm\delta$ are the corresponding NN, NNN and TNN lattice vectors as shown in Supplemental Fig. 1(a) \cite{supplement}, and $t_{\zeta,\zeta^\prime}$, $r_{\zeta,\zeta^\prime}$ and $u_{\zeta,\zeta^\prime}$ are the  corresponding hoppings. In general, the disorder effects on low-dimensional materials include two types. One is the local change of the on-site energy, and the other is the changes of the electronic hoppings, owing to the changes of bond lengths and angles \cite{AHCastroNeto2009}. A typical example of the second type is the lattice deformation, where the strain enters into the Hamiltonian in the form of both scalar potential and effective magnetic vector potential. For the clamped ion models, it is assumed that the original lattice remains but the electronic hoppings are mainly modified by the changes of bond lengths \cite{polarization1,chernDroth,chernLiWang,chernRostami,AHCastroNeto2009}. In this way, the strain-modified hopping terms in the linear elastic approximation are written as $J_{\bm r,\zeta-\zeta^\prime}(\bm\varepsilon)=J^0_{\bm r,\zeta-\zeta^\prime}[1-(\beta_{ij}/|\bm {r}|^2)\sum_{mn} r_m\varepsilon_{mn} r_n]$, where $\bm {r}=\bm{R}_j-\bm{R}_i$ is the lattice vectors pointed from $i$ to $j$ in Supplemental Fig. 1(a) \cite{supplement}, $\bm\varepsilon$ is the strain, $m$ and $n$ denote $x$ or $y$, $J^0_{\bm r}$ denotes the initial hoppings for undeformed TMDMs in Supplemental Table I \cite{supplement}, and the dimensionless EGP is defined as $\beta_{ij}=-[d\ln J_{\bm r}(r^s)/d\ln r^s]|_{\bm{\varepsilon}\rightarrow \bm{0}}$ to measure how the electronic hopping changes with the bond length $(r^s=|\bm {r}^s(\bm\varepsilon)|)$. Because the dominant hopping orbitals arise from all the $d$ orbitals of M atoms \cite{dorbitals1}, it is reasonable to approximately assume $\beta_{ij}$ as a constant $\beta$. Using the Fourier transforms, we obtain the strain-modulated Hamiltonian $H(\bm k,\bm\varepsilon)$ in momentum space (Supplemental Materials \cite{supplement}).
\begin{figure*}
\begin{center}
\includegraphics[width=15.76cm,angle=0]{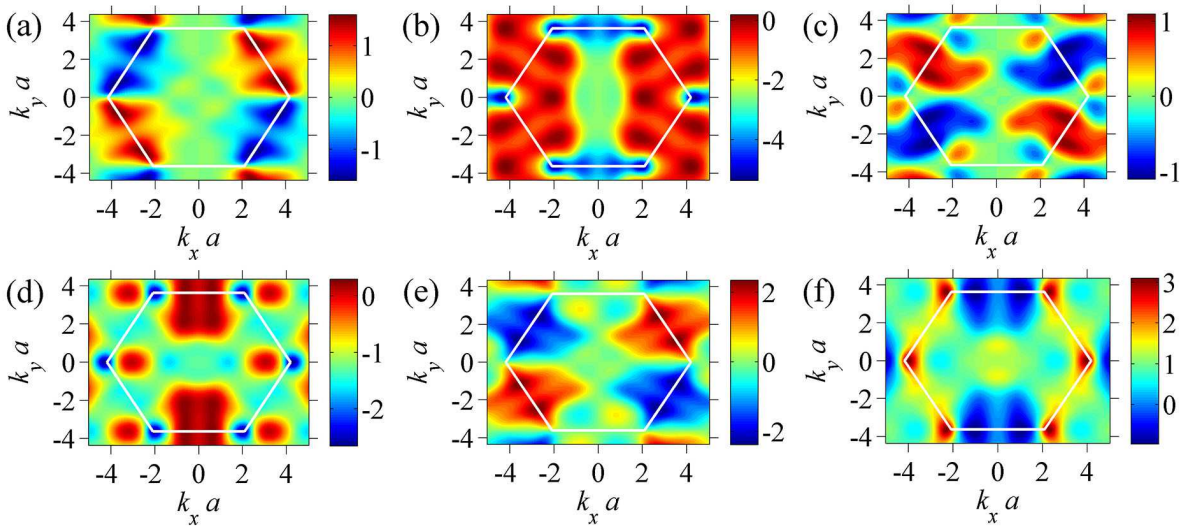}
\caption{\label{fig1} The contour maps of piezoelectric Berry curvature (in units of {\AA}) as a function of $k_xa$ and $k_ya$ in momentum space for MoS$_2$: (a) $\Omega_{1,11}$, (b) $\Omega_{1,12}$, (c) $\Omega_{1,22}$, (d) $\Omega_{2,11}$, (e) $\Omega_{2,12}$, and (f) $\Omega_{2,22}$. The white hexagons show the BZ.}
\end{center}
\end{figure*}
In the first order approximation of the atomic displacements, the electron-phonon coupling Hamiltonian $H_{ep}$ reads \cite{FGiustino2017}
\begin{equation}
  H_{ep}=\frac{1}{\sqrt{A}}\sum_{\bm k,p;\bm k^\prime,p^\prime;\bm q,\nu}g^\nu_{\bm k,p;\bm k^\prime,p^\prime}c^\dag_{\bm k,p}c_{\bm k^\prime,p^\prime}\phi_{\bm q,\nu},
\end{equation}
where A is the area, $\phi_{\bm q,\nu}=\sqrt{\hbar\omega_{\bm q,\nu}/2}(b_{\bm q,\nu}+b^\dag_{-\bm q,\nu})$ is the phonon operator with its wave vector $\bm q$, $\bm k$ and $\bm k^\prime$ are the electron wave vectors, $p$ and $p^\prime$ are the energy band indexes. The electron-phonon matrix element $g^\nu_{\bm k,p;\bm k^\prime,p^\prime}$ in the TB approximation reads
\begin{equation}
 g^\nu_{\bm k,p;\bm k^\prime,p^\prime}=\frac{\sum_{n^\prime,s,s^\prime}U^*_{s,p}(\bm k)U_{s^\prime,p^\prime}(\bm k^\prime)\bm O\cdot\bm\nabla J(0,s;n^\prime,s^\prime)|_0}{\sqrt{\mu} \omega_{\bm k-\bm {k}^\prime,\nu}},
 \label{eqEPcoupling}
\end{equation}
where $\mu$ is the mass per unit area, $U$ is the unitary matrix that diagonalizes the strain-independent $H(\bm k,\bm\varepsilon)|_{\bm{\varepsilon}\rightarrow \bm{0}}$, $\bm O=e^{i\bm k^\prime\cdot\bm R_{n^\prime}}\bm S^\nu_s(\bm k -\bm k^\prime)-e^{i\bm k\cdot\bm R_n}\bm S^\nu_{s^\prime}(\bm k -\bm k^\prime)$ with the phonon polarization vector $\bm S^\nu_s$, and $J(0,s;n^\prime,s^\prime)$ is the matrix element of the NN interactions between atoms $(0,s)$ and $(n^\prime,s^\prime)$ \cite{TBeph}. Using $\beta=-[d\ln J_{\bm r}(r^s)/d\ln r^s]|_{\bm{\varepsilon}\rightarrow \bm{0}}$, we have $\bm\nabla J(0,s;n^\prime,s^\prime)|_0\cdot\bm\hat{r}=\beta J^0_{\bm r}/r$, with the NN unit vector $\bm\hat{r}$ and length $r$ of $\bm r$. Owing to $|g^\nu_{\bm k,p;\bm k^\prime,p^\prime}|\propto\beta$, consequently, EGP is usually used to characterize the electron-phonon coupling within the TB framework \cite{ValleyPseuMagnetic1,ValleyPseuMagnetic2,ValleyPseuMagnetic3, ValleyPseuMagneticGphe,TBeph}.

\textit{Electronic contribution to piezoelectric coefficients.} Under a time-dependent non-electromagnetic perturbation with a slow variation, the semiclassical dynamical equation for the $n$-th energy band reads
$\dot{\bm{r}}_n=[\partial E^0_n(\bm{k})/\partial (\hbar\bm{k})]-\Omega^n_{\bm{k},t}$, and $\dot{\bm{k}}=0$, where $E^0_n(\bm{k})$ is the energy for the system without perturbations, and $\Omega^n_{\bm{k},t}$ is the Berry curvature with respect to $\bm{k}$ and $t$. Therefore, the perturbation-induced adiabatic charge current in 2D systems reads $\bm{j}(t)=2e\sum_n \int_{BZ}\Omega^n_ {\bm{k},t} d\bm{k}/(2\pi)^2$.  The piezoelectric adiabatic process varies from the initial state with $\bm{\varepsilon}(0)=\bm{0}$ to the final state with $\bm{\varepsilon}(T)= \bm{\varepsilon}$, \emph{i.e.}, $\bm{\varepsilon}(T)= \Delta\bm{\varepsilon}$, owing to $\bm{\varepsilon}-\bm{0}= \Delta\bm{\varepsilon}$, and hence the piezoelectric adiabatic current is $\bm{j}(t)=2e\sum_{jk}\sum_n \int_{BZ}\Omega^n_ {\bm{k},\varepsilon_{jk}}\dot{\varepsilon}_{jk} d\bm{k}/(2\pi)^2$. The continuity equation and the relation between polarization and charge densities require the polarization difference along the $i$ direction to satisfy $\Delta P_i=\int^T_0 j_i(t)dt=\sum_{jk}\left(2e\sum_n\int_{BZ} \Omega^n_{k_i,\varepsilon_{jk}}d\bm{k}/(2\pi)^2\right)
\Delta\varepsilon_{jk}$. Then the electronic part of piezoelectric coefficient $e_{ijk}$ for TMDMs, \emph{i.e.}, $e_{ijk} = (\partial P_i/\partial \varepsilon_{jk})|_{\bm{\varepsilon}\rightarrow \bm{0}}$, reads
\begin{subequations}
\label{pizocoef}
\begin{eqnarray}
e_{ijk}=\frac{e}{2\pi^2}\int_{BZ}\Omega_{i,jk}d\bm{k},\label{pizocoef1}
\end{eqnarray}
\begin{equation}
  \Omega_{i,jk}=i\sum^2_{m=1}\frac{\left\langle u^0_v|v_i|u^0_{c_m}\right\rangle \left\langle u^0_{c_m}|w_{jk}|u^0_v\right\rangle-c.c.}{(E^0_v-E^0_{c_m})^2},\label{pizocoef2}
\end{equation}
\end{subequations}
where $\Omega_{i,jk}$ is a short-hand notation of $\Omega_{k_i,\varepsilon_{jk}}|_{\bm{\varepsilon}\rightarrow \bm{0}}$, $E^0_v(E^0_c)$ and $u^0_v(u^0_c)$ are the corresponding eigenvalues and normalized eigenstates with one valance ($v$) band and two conduction ($c$) bands for the unstrained TMDMs, as shown in Supplemental Fig. 1(b) \cite{supplement}, $v_i=\partial [H(\bm k,\bm\varepsilon)|_{\bm{\varepsilon}\rightarrow \bm{0}}]/\partial k_i$, $w_{jk}=\partial H(\bm k,\bm\varepsilon) / \partial \varepsilon_{jk}$, and \emph{c.c.} denotes the complex conjugate.
Let us use the TB piezoelectric model to calculate both the piezoelectric Berry curvature and coefficients for MoS$_2$, as an example of TMDMs because of their similarity. We adopt the TB parameters in Table III of Ref. \cite{dorbitals1} and $\beta=2$ \cite{EPGbeta}. Considering $\varepsilon_{12}=\varepsilon_{21}$, we have $e_{112} = e_{121}$, and $e_{212} = e_{221}$. Therefore, only six piezoelectric coefficients need to be calculated. Figures ~\ref{fig1} show the distributions of the Berry curvature $\Omega_{i,jk}$ for MoS$_2$ in momentum space. It can be seen that, $\Omega_{1,11}$, $\Omega_{1,22}$ and $\Omega_{2,12}$ are odd functions of $k_x$ and $k_y$ inside the BZ in Figs.~\ref{fig1}(a), ~\ref{fig1}(c) and ~\ref{fig1}(e), respectively. Consequently, corresponding piezoelectric coefficients, $e_{111}$, $e_{122}$, $e_{212}$ and $e_{221}$, as their integrals over the BZ, must be zero. The other $\Omega_{1,12}$, $\Omega_{2,11}$ and $\Omega_{2,22}$ are even functions of $k_x$ and $k_y$ inside the BZ in Figs.~\ref{fig1}(b), ~\ref{fig1}(d) and ~\ref{fig1}(f), respectively, and hence $e_{112}$, $e_{211}$ and $e_{222}$ are nonzero. The values of these piezoelectric coefficients are listed in Table~\ref{tab1}. In general, the $D_{3h}$ point group requires $e_{111} = e_{122} = e_{212} = e_{221}=0$, and $e_{211} = -e_{222} = e_{112}/2 = e_{121}/2$, if the piezoelectric coefficient is defined as $e_{ijk} = (\partial P_i/\partial \varepsilon_{jk})|_{\bm{\varepsilon}\rightarrow \bm{0}}$ \cite{chernLiWang}. The results in Table~\ref{tab1} obey the symmetry of $D_{3h}$ and agree well with both DFT \cite{piezoFP2dduerloo} and experiments \cite{piezoExpTMDCXZhang}. It should be figured out that the experimental result is corresponding to the relaxed-ion piezoelectric coefficient contributed by both electrons and ions. In Table~\ref{tab1}, the clamped-ion TB result  is approximately comparable to the experiment, because the electronic contribution is much larger than the ionic part for MoS$_2$ and WS$_2$ \cite{piezoFP2dduerloo,piezolattdynamical}, but note that the comparison for other TMDMs is inappropriate.
\begin{table}[b]
\caption{\label{tab1}
The obtained piezoelectric coefficients $(10^{-10}C/m)$ of MoS$_2$ from the TB piezoelectric theory. Exp. denotes the experimental values.}
\begin{ruledtabular}
\begin{tabular}{ccccccc}
 & $e_{111}$ & $e_{122}$ & $e_{212}$ & $e_{112}$ & $e_{211}$ & $e_{222}$\\
\hline
TB & 0 & 0 & 0 & $-$5.83 & $-$2.89 & 2.87 \\
DFT \cite{piezoFP2dduerloo} & - & - & - & - & - & 3.06 \\
Exp. \cite{piezoExpTMDCXZhang} & - & - & - & - & - & {2.9$\pm0.5$} \\
\end{tabular}
\end{ruledtabular}
\end{table}
\begin{table}[b]
\caption{\label{tab2}
EGP $(\beta)$ for TMDMs. $e^0_{222}$ ($10^{-10}C/m$) is a calculated reference value for $\beta_0=1$ by Eqs.~(\ref{pizocoef}).}
\begin{ruledtabular}
\begin{tabular}{ccccccc}
 & MoS$_2$ & MoSe$_2$ & MoTe$_2$ & WS$_2$ & WSe$_2$ & WTe$_2$ \\
\hline
$e_{222}$ \cite{piezoFP2dduerloo} & 3.06 & 2.80 & 2.98 & 2.20 & 1.93 & 1.60 \\
$e^0_{222}$ & 1.4365 & 1.4057 & 1.3672 & 1.4450 & 1.4560 & 1.4002 \\
$\beta=e_{222}/e^0_{222}$ & 2.13 & 1.99 & 2.18 & 1.52 & 1.33 & 1.14 \\
\end{tabular}
\end{ruledtabular}
\end{table}

\textit{EGP of TMDMs.} EGP is important to determine the electron-phonon coupling, PMF \cite{ValleyPseuMagnetic1,ValleyPseuMagnetic2, ValleyPseuMagnetic3,ValleyPseuMagnetic4} and strain-modulated electronic transports in 2D materials \cite{AHCastroNeto2009,EPGTransp2D}. Therefore, it is meaningful to predict EGP of TMDMs. The strain-modulated $H(\bm k,\bm\varepsilon)$ and Eqs.~(\ref{pizocoef}) show that $e_{ijk}$ is proportional to EGP $(\beta)$, \emph{i.e.}, $e_{ijk}\propto\beta$. Therefore, if the clamped-ion piezoelectric coefficient has been obtained from DFT, one can inversely determine EGP, \emph{i.e.}, $\beta/\beta_0=e_{ijk}/e^0_{ijk}$, where $e^0_{ijk}$ is a \emph{calculated reference value} of the TB piezoelectric coefficient for $\beta_0=1$ by Eq.~(\ref{pizocoef}), according to the consistence between the TB piezoelectric model and DFT results. The obtained results in Table~\ref{tab2} show that $\beta$ for Mo-based TMDMs is about 2. In addition, $\beta$ for Mo-based TMDMs is larger than that for W-based TMDMs. It means that the electronic hoppings for Mo $4d$ orbitals change faster with the bond length than that for W $5d$ orbitals. Physically, because the distribution of W $5d$ orbitals in space is wider than that of Mo $4d$ orbitals, the interaction among W $5d$ orbitals is more robust against the strain than that of Mo $4d$ orbitals.

\textit{Analytical piezoelectric model and PMF.} As shown in Fig.~\ref{fig1}(f), the Berry curvature $\Omega_{2,22}$ is mainly located in the six corners of the BZ. This means that $e_{222}$ is mainly contributed by the Berry curvatures in the vicinity of $K(K^\prime)$ point. Therefore, it is effective to analytically evaluate $e_{222}$ by the strain-dependent $\bm{k\cdot p}$ Hamiltonian \cite {supplement} and Eqs.~(\ref{pizocoef}). Because the SOC splitting is much less than the band gap in TMDMs \cite{dorbitals1,dorbitals2}, SOC has a weak influence on the piezoelectricity. In addition, the time-reversal invariant allows us to consider the piezoelectricity only at the $K$ valley, because of the valley degenerate. We first consider the piezoelectricity based on the first order $\bm{k\cdot p}$ Hamiltonian. The trigonal warping effects on the piezoelectricity are also explored in Supplemental Materials \cite{supplement}. For the first order approximation, the Berry curvatures $\Omega^s_{2,22}$ and $\Omega^v_{2,22}$ induced by the corresponding strain-induced scalar and vector potentials read (Supplemental Materials \cite{supplement})
\begin{subequations}
\label{BerryCA}
\begin{eqnarray}
\Omega^s_{2,22}=\frac{a^2t^2(\alpha_c-\alpha_v)\beta q_x}{4[a^2t^2(q^2_x+q^2_y)+(\Delta/2)^2]^{3/2}},\label{BerryCAs}
\end{eqnarray}
\begin{equation}
  \Omega^v_{2,22}=\frac{at\alpha \beta \Delta}{4[a^2t^2(q^2_x+q^2_y)+(\Delta/2)^2]^{3/2}},\label{BerryCAv}
\end{equation}
\end{subequations}
where $\alpha_c$ and $\alpha_v$ are the energy parameters of conduction and valence bands for the scalar potential, respectively, and $\alpha$ is the energy parameters for the vector potential. Then the piezoelectric coefficient reads
\begin{eqnarray}
e_{222} &=& \frac{4e}{(2\pi)^2} \int_0^{2\pi} d\theta \int_0^{q_m}(\Omega^s_{2,22}+\Omega^v_{2,22})q dq \nonumber \\
&=&\frac{e\alpha\beta}{\pi at} \left( 1-\frac{\sqrt{3}\Delta}{\sqrt{16\sqrt{3}\pi t^2+3\Delta^2}}\right),\label{pizoeff}
\end{eqnarray}
where the factor $4$ contains both the spin and valley degenerates, $q = \sqrt{q_x^2+q_y^2}$, and $\pi q_m^2 = S_{BZ}/2$ with the first BZ area $S_{BZ}$ $( 8\pi^2/{\sqrt{3}a^2})$. Equation ~(\ref{BerryCAs}) shows that $\Omega^s_{2,22}$ is an odd function of $q_x$. Therefore, $\Omega^s_{2,22}$ affects the work function \cite{strainWorkFuc} but has no contributions to piezoelectricity, because its integral is zero. Consequently, $\Omega^v_{2,22}$ mainly contributes to $e_{222}$.  The analytical expression of $e_{222}$ in Eq.~(\ref{pizoeff}) demonstrates the clear relation among the piezoelectricity, lattice constant, band gap, EGP and energy parameters ($t$ and $\alpha$), and hence it provides a direct estimation of the piezoelectric coefficient for TMDMs.
\begin{table}[b]
\caption{\label{tab3}
PMF strength $B_0$ ($10^{-5}$T$\cdot$m) for TMDMs. The energy parameters $t$ and $\Delta$ (in units of eV) are obtained from the fitting between the first order continuum approximation and tight-binding model.}
\begin{ruledtabular}
\begin{tabular}{ccccccc}
 & MoS$_2$ & MoSe$_2$ & MoTe$_2$ & WS$_2$ & WSe$_2$ & WTe$_2$ \\
\hline
$t$ & 1.0799 & 0.8984 & 0.7213 & 1.3315 & 1.1395 & 0.9739 \\
$\Delta$ & 1.6579 & 1.4293 & 1.2302 & 1.8062 & 1.5412 & 1.0668 \\
$B_0$ & 5.4416 & 5.0429 & 5.51 & 3.7576 & 3.2935 & 2.5795 \\
\end{tabular}
\end{ruledtabular}
\end{table}
For a deformation without strain gradients, there is no strain-induced PMF in 2D materials. In addition, the lattice correction has no contributions to PMF, and hence PMF only contains the electronic parts. In general, if the armchair direction of TMDMs has an angle $\theta$ to the $x$ axis, the PMF, $B = (\nabla\times\bm A)\cdot\bm{\bm e}_z$, is written as \cite{supplement}
\begin{eqnarray}
  \frac{B}{B_0}=-\left[\frac{\partial(\varepsilon_{xx}-\varepsilon_{yy})}{\partial x}-2\frac{\partial \varepsilon_{xy}}{\partial y}\right]\sin(3\theta) \nonumber \\ -\left[\frac{\partial(\varepsilon_{xx}-\varepsilon_{yy})}{\partial y}+2\frac{\partial \varepsilon_{xy}}{\partial x}\right]\cos(3\theta),
\end{eqnarray}
where $B_0=\hbar\alpha\beta/aet$ represents the PMF strength. Using Eq.~(\ref{pizoeff}) and $B_0=\hbar\alpha\beta/aet$, we cancel the unknown factor $\alpha\beta$ and further write the PMF strength as
\begin{equation}
  B_0=\frac{\pi\hbar\sqrt{16\sqrt{3}\pi t^2+3\triangle^2}}{e^2\left(\sqrt{16\sqrt{3}\pi t^2+3\triangle^2}-\sqrt{3}\triangle\right)}e_{222}.\label{pseuMF}
\end{equation}
The calculated $B_0$ for TMDMs by Eq.~(\ref{pseuMF}) is listed in Table~\ref{tab3}. For the same deformation $B_0$ for TMDMs is larger than that for graphene with its PMF strength $(B_0\sim 1\times10^{-5}$T$\cdot$m) \cite{ValleyPseuMagneticGphe}. The strain-induced large PMF in TMDMs is manifested by the giant valley drift \cite{GiantVallshif}. Therefore, strain even with a small magnitude has a remarkable effect on the electronic, optical and magnetic properties of TMDMs \cite{strainElectron,strainOptEle,strainWorkFuc,epValleyMagnet}.

\textit{Discussion.} We now briefly comment on the experiments for the predicted EGP and PMF. Although the phonon Gr\"{u}neisen parameter can be extracted from the Raman spectroscopy \cite{strainEffphonon1,RammanExp}, it is challenging to experimentally measure EGP $(\beta)$. Recently, the time- and angle-resolved photoemission spectroscopy (tr-ARPES) has been extended to map momentum-resolved electronic structure and electron-phonon interaction \cite{ARPESZXShen}. Therefore, the strain-tunable electronic band structure and electron-phonon coupling \cite{epExp} for TMDMs can be mapped by tr-ARPES such that $\beta$ is determined, similar to that of carbon nanotube \cite{Cnanotube}. The strain gradient induced by the nanobubbles \cite{bubble} or a uniaxial stretch \cite{stretch} induces PMF, which leads to pseudomagnetic quantum Hall effect \cite{QHE}. Consequently, the scanning tunneling microscopy and spectroscopy are used to measure the PMF \cite{STM}.

\textit{Conclusion.} We develop an atomic-scale TB piezoelectric theory for TMDMs. By means of the consistence between the TB and clamped-ion DFT piezoelectric models, we predict the values of EGP for TMDMs. Using the TB piezoelectric model in the continuum approximation, we further build the electromechanical coupling correlation between the electronic part of piezoelectric coefficient and strain-induced PMF and also forecast the values of PMF for TMDMs. Our results will spark more interest in electromechanical couplings in 2D materials and benefit novel atomically-thin piezotronics and straintronics.

This work was supported financially by National Natural Science Foundation of China under Grant Nos. 11502308, 11472313, 11232015 and 11572355, Guangdong Natural Science Foundation of China under Grant No. 2016A030310205, and the fundamental research funds for the central universities under Grant No. 17lgpy31.

\textit{Note added.} We note two recent studies \cite{chernRostami,Addnote}. In Ref. \cite{chernRostami}, they obtain the piezoelectric coefficients for some of TMDMs in the continuum approximation. In Ref. \cite{Addnote}, the TB Hamiltonian for 2D strained hexagonal crystals is also obtained.

\end{document}


\title{Supplemental Materials: Tight-binding piezoelectric theory and electromechanical coupling correlations for transition metal dichalcogenide monolayers}

\author{Yunhua Wang}
\email{wangyh49@mail.sysu.edu.cn}
\affiliation{Sino-French Institute of Nuclear Engineering and Technology, Sun Yat-sen University, Zhuhai 519082, China}

\author{Zongtan Wang}
\thanks{Z. W. contributed equally to this work as Y. H. W..}
\affiliation{School of Engineering, Sun Yat-sen University, Guangzhou 510006, China}

\author{Jie Li}
\affiliation{School of Engineering, Sun Yat-sen University, Guangzhou 510006, China}

\author{Jie Tan}
\affiliation{Sino-French Institute of Nuclear Engineering and Technology, Sun Yat-sen University, Zhuhai 519082, China}

\author{Biao Wang}
\email{wangbiao@mail.sysu.edu.cn}
\affiliation{Sino-French Institute of Nuclear Engineering and Technology, Sun Yat-sen University, Zhuhai 519082, China}
\affiliation{State Key Laboratory of Optoelectronic Materials and Technologies, School of Physics, Sun Yat-sen University, Guangzhou 510275, China}

\author{Yulan Liu}
\email{stslyl@mail.sysu.edu.cn}
\affiliation{School of Engineering, Sun Yat-sen University, Guangzhou 510006, China}


\pacs{Valid PACS appear here}
\maketitle

\section{\label{sec1}Strain-dependent TB Hamiltonian in momentum space}
\begin{table*}[b]
\caption{\label{Stab1}
Hoppings between different $d$ orbitals of M atoms for MX$_2$. The first row shows the combinations of arbitrary two $d$ orbitals, the first column displays the NN, NNN and TNN lattice vectors, and the other columns show their corresponding hoppings for different lattice vectors. The values of these TB parameters are listed in Table III of Ref. [47] of the main text.}
\begin{ruledtabular}
\begin{tabular}{ccccccc}
 & $d_{z^2}-d_{z^2}$ & $d_{xy}-d_{xy}$ & $d_{x^2-y^2}-d_{x^2-y^2}$ & $d_{z^2}-d_{xy}$ & $d_{z^2}-d_{x^2-y^2}$ & $d_{xy}-d_{x^2-y^2}$\\
\hline
$\bm{\delta_1}$ & $t_0$ & $t_{11}$ & $t_{22}$ & $t_1$ & $t_2$ & $t_{12}$ \\
$\bm{\delta_2}$ & $t_0$ & $\frac{t_{11}+3t_{22}}{4}$ & $\frac{3t_{11}+t_{22}}{4}$ & $\frac{t_1-\sqrt{3}t_2}{2}$ & $-\frac{t_2+\sqrt{3}t_1}{2}$ & $\frac{\sqrt{3}(t_{22}-t_{11})}{4}-t_{12}$ \\
$\bm{\delta_3}$ & $t_0$ & $\frac{t_{11}+3t_{22}}{4}$ & $\frac{3t_{11}+t_{22}}{4}$ & $-\frac{t_1-\sqrt{3}t_2}{2}$ & $-\frac{t_2+\sqrt{3}t_1}{2}$ & $\frac{\sqrt{3}(t_{11}-t_{22})}{4}+t_{12}$ \\
$\bm{\delta_4}$ & $t_0$ & $t_{11}$ & $t_{22}$ & $-t_1$ & $t_2$ & $-t_{12}$ \\
$\bm{\delta_5}$ & $t_0$ & $\frac{t_{11}+3t_{22}}{4}$ & $\frac{3t_{11}+t_{22}}{4}$ & $-\frac{t_1+\sqrt{3}t_2}{2}$ & $-\frac{t_2-\sqrt{3}t_1}{2}$ & $\frac{\sqrt{3}(t_{22}-t_{11})}{4}+t_{12}$ \\
$\bm{\delta_6}$ & $t_0$ & $\frac{t_{11}+3t_{22}}{4}$ & $\frac{3t_{11}+t_{22}}{4}$ & $\frac{t_1+\sqrt{3}t_2}{2}$ & $-\frac{t_2-\sqrt{3}t_1}{2}$ & $\frac{\sqrt{3}(t_{11}-t_{22})}{4}-t_{12}$ \\
$\bm{\chi_1}$ & $r_0$ & $r_{11}$ & $\frac{\sqrt{3}r_{11}+2r_{12}}{\sqrt{3}}$ & $r_1$ & $-\frac{r_1}{\sqrt{3}}$ & $r_{12}$ \\
$\bm{\chi_2}$ & $r_0$ & $r_{11}+\sqrt{3}r_{12}$ & $\frac{\sqrt{3}r_{11}-r_{12}}{\sqrt{3}}$ & $0$ & $\frac{2r_2}{\sqrt{3}}$ & $0$ \\
$\bm{\chi_3}$ & $r_0$ & $r_{11}$ & $\frac{\sqrt{3}r_{11}+2r_{12}}{\sqrt{3}}$ & $-r_1$ & $-\frac{r_1}{\sqrt{3}}$ & $-r_{12}$ \\
$\bm{\chi_4}$ & $r_0$ & $r_{11}$ & $\frac{\sqrt{3}r_{11}+2r_{12}}{\sqrt{3}}$ & $r_2$ & $-\frac{r_2}{\sqrt{3}}$ & $r_{12}$ \\
$\bm{\chi_5}$ & $r_0$ & $r_{11}+\sqrt{3}r_{12}$ & $\frac{\sqrt{3}r_{11}-r_{12}}{\sqrt{3}}$ & $0$ & $\frac{2r_1}{\sqrt{3}}$ & $0$ \\
$\bm{\chi_6}$ & $r_0$ & $r_{11}$ & $\frac{\sqrt{3}r_{11}+2r_{12}}{\sqrt{3}}$ & $-r_2$ & $-\frac{r_2}{\sqrt{3}}$ & $-r_{12}$ \\
$2\bm{\delta_1}$ & $u_0$ & $u_{11}$ & $u_{22}$ & $u_1$ & $u_2$ & $u_{12}$ \\
$2\bm{\delta_2}$ & $u_0$ & $\frac{u_{11}+3u_{22}}{4}$ & $\frac{3u_{11}+u_{22}}{4}$ & $\frac{u_1-\sqrt{3}u_2}{2}$ & $-\frac{u_2+\sqrt{3}u_1}{2}$ & $\frac{\sqrt{3}(u_{22}-u_{11})}{4}-u_{12}$ \\
$2\bm{\delta_3}$ & $u_0$ & $\frac{u_{11}+3ut_{22}}{4}$ & $\frac{3u_{11}+u_{22}}{4}$ & $-\frac{u_1-\sqrt{3}u_2}{2}$ & $-\frac{u_2+\sqrt{3}u_1}{2}$ & $\frac{\sqrt{3}(u_{11}-u_{22})}{4}+u_{12}$ \\
$2\bm{\delta_4}$ & $u_0$ & $u_{11}$ & $u_{22}$ & $-u_1$ & $u_2$ & $-u_{12}$ \\
$2\bm{\delta_5}$ & $u_0$ & $\frac{u_{11}+3u_{22}}{4}$ & $\frac{3u_{11}+u_{22}}{4}$ & $-\frac{u_1+\sqrt{3}u_2}{2}$ & $-\frac{u_2-\sqrt{3}u_1}{2}$ & $\frac{\sqrt{3}(u_{22}-u_{11})}{4}+u_{12}$ \\
$2\bm{\delta_6}$ & $u_0$ & $\frac{u_{11}+3u_{22}}{4}$ & $\frac{3u_{11}+u_{22}}{4}$ & $\frac{u_1+\sqrt{3}u_2}{2}$ & $-\frac{u_2-\sqrt{3}u_1}{2}$ & $\frac{\sqrt{3}(u_{11}-u_{22})}{4}-u_{12}$ \\
\end{tabular}
\end{ruledtabular}
\end{table*}
\begin{figure}
\begin{center}
\includegraphics[width=8.6cm,angle=0]{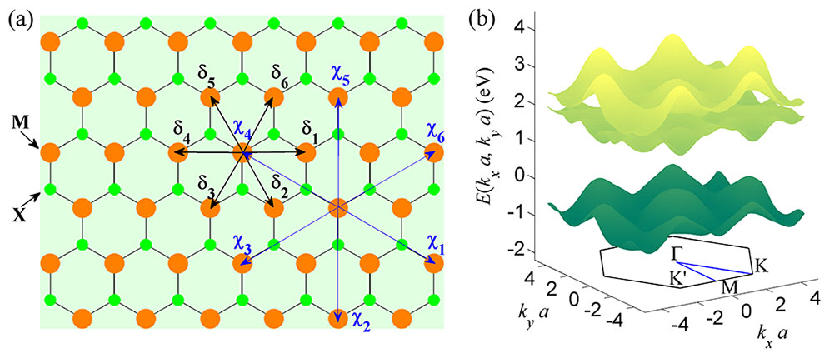}
\caption{\label{Sfig1} (a) A top view of the trigonal prismatic structure for MX$_2$ with the lattice constant $a$ and the corresponding NN, NNN and TNN lattice vectors $\bm\delta_i$, $\bm\chi_i$ and $2\bm\delta_i$. (b) The three-band TB energy dispersion for MoS$_2$ with one negative valence band and two positive conduction bands, where the top of valence band is shifted to zero, and the hexagon shows the first Brillouin zone (BZ).}
\end{center}
\end{figure}
Figure~\ref{Sfig1}(a) shows the top view of the lattice structure for TMDMs (MX$_2$), where the nearest-neighbor (NN), next-nearest-neighbor (NNN) and third-nearest-neighbor (TNN) lattice vectors $\bm r$ are denoted by  $\bm\delta_i$, $\bm\chi_i$ and $2\bm\delta_i$, respectively, with $i=1,\cdots,6$. In general, the strain is defined as $\varepsilon_{mn}=(1/2)[\partial u^2_z/\partial r_m\partial r_n+\partial u_m/\partial r_n+\partial u_n/\partial r_m]$, with the displacement vector $\bm u$. We consider the in-plain intrinsic piezoelectricity, where the displacement component $u_z$ is zero. Consequently, the plain strain tensor is further written as $\varepsilon_{mn}=(1/2)[\partial u_m/\partial r_n+\partial u_n/\partial r_m]$. Using $J_{\bm r,\zeta-\zeta^\prime}(\bm\varepsilon)=J^0_{\bm r,\zeta-\zeta^\prime}[1-(\beta/|\bm {r}|^2)\sum_{mn} r_m\varepsilon_{mn} r_n]$ within the clamped ion framework, we can rewrite Eq. (1) of the main text as the strain-modulated Hamiltonian. Via the following Fourier transforms,
\begin{equation}
 d_{i,\zeta}=\frac{1}{N}\sum_{\bm k} e^{i\bm{k}\cdot\bm{R_i}}d_{\bm k,\zeta}, d^\dag_{i,\zeta}=\frac{1}{N}\sum_{\bm k} e^{-i\bm{k}\cdot\bm{R_i}}d^\dag_{\bm k,\zeta},
\end{equation}
we further write the strain-dependent TB Hamiltonian in momentum space as
\begin{equation}
 H(\bm \varepsilon)=\sum_{\bm k}
\left[
\begin{array}{ccc}
d^\dag_{\bm k,z^2} & d^\dag_{\bm k,xy} & d^\dag_{\bm k,x^2-y^2}
\end{array}\right]H(\bm k,\bm \varepsilon)
\left[
\begin{array}{c}
d_{\bm k,z^2}\\
d_{\bm k,xy}\\
d_{\bm k,x^2-y^2}
\end{array}\right].
\end{equation}
The Hamiltonian $H(\bm k,\bm \varepsilon)$ consists of the NN $(\bm \delta)$, NNN $(\bm\chi)$ and TNN $(2\bm\delta)$ components, \emph{i.e.},
\begin{equation}
\label{eq3}
H(\bm k,\bm \varepsilon)=H_{\bm \delta}(\bm k,\bm \varepsilon)+H_{\bm \chi}(\bm k,\bm \varepsilon)+H_{2\bm \delta}(\bm k,\bm \varepsilon).
\end{equation}
The matrix elements for $H^{i,j}_{\bm \delta}(\bm k,\bm \varepsilon)$, $H^{i,j}_{\bm \chi}(\bm k,\bm \varepsilon)$ and $H^{i,j}_{2\bm \delta}(\bm k,\bm \varepsilon)$ with $(i=1,2,3;j\geq i)$ read
\begin{subequations}
\label{eq4}
\begin{equation}
 H^{i,j}_{\bm\delta}(\bm k,\bm \varepsilon)= \sum^6_{l=1}J_{\bm{\delta_l},\zeta_i-\zeta_j}(\bm\varepsilon)e^{i\bm k\cdot \bm{\delta_l}}+\epsilon_i\delta_{ij},
\end{equation}
\begin{equation}
 H^{i,j}_{\bm\chi}(\bm k,\bm \varepsilon)= \sum^6_{l=1}J_{\bm{\chi_l},\zeta_i-\zeta_j}(\bm\varepsilon)e^{i\bm k\cdot \bm{\chi_l}},
\end{equation}
\begin{equation}
 H^{i,j}_{2\bm\delta}(\bm k,\bm \varepsilon)=
 \sum^6_{l=1}J_{2\bm{\delta_l},\zeta_i-\zeta_j}(\bm\varepsilon)e^{i\bm k\cdot 2\bm{\delta_l}},
\end{equation}
\end{subequations}
where $\epsilon_1$, $\epsilon_2$ and $\epsilon_3$ ($\epsilon_3=\epsilon_2$) are the on-site energy, $\delta_{ij}$ is the Kronecker delta, $\zeta_i$ and $\zeta_j$ denote the $d$ orbitals in Table~\ref{Stab1}. The obtained strain-dependent Hamiltonian $H(\bm k,\bm \varepsilon)$ in Eqs. (\ref{eq3}) and (\ref{eq4}) can be used to calculate the piezoelectric Berry curvature $\Omega_{i,jk}$ such that the electronic part of piezoelectric coefficients can be further evaluated.
\section{\label{sec2}Strain-dependent $\bm{k\cdot p}$ Hamiltonian and pseudomagnetic vector potential}
Expanding the three-band Hamiltonian $H(\bm k,\bm \varepsilon)$ at the K valley and using L\"{o}wdin partitioning method \cite{Lowdin}, we write the strain-dependent $\bm{k\cdot p}$ Hamiltonian as
\begin{subequations}
\label{ContHami}
\begin{eqnarray}
 H_K(\bm q,\bm \varepsilon)=H^{1}_{0}(\bm q)+H^{2}_{0}(\bm q)+H^{s}_{S}(\bm \varepsilon)+H^{v}_{S}(\bm \varepsilon),
\end{eqnarray}
\begin{equation}
H^{1}_{0}(\bm q)=
\left[
\begin{array}{cc}
\Delta/2 & at(q_x-iq_y) \\
at(q_x+iq_y) & -\Delta/2
\end{array}\right],
\end{equation}
\begin{equation}
H^{2}_{0}(\bm q)=
\left[
\begin{array}{cc}
\gamma_1a^2(q^2_x+q^2_y) & \gamma_3a^2(q_x+iq_y)^2 \\
\gamma_3a^2(q_x-iq_y)^2 & \gamma_2a^2(q^2_x+q^2_y)
\end{array}\right],
\end{equation}
\begin{equation}
H^{s}_{S}(\bm \varepsilon)=
\left[
\begin{array}{cc}
\alpha_c\beta(\varepsilon_{xx}+\varepsilon_{yy}) & 0 \\
0 & \alpha_v\beta(\varepsilon_{xx}+\varepsilon_{yy})
\end{array}\right],
\end{equation}
\begin{equation}
H^{v}_{S}(\bm \varepsilon)=
\alpha\beta\left[
\begin{array}{cc}
0 & \varepsilon_{xx}-\varepsilon_{yy}-i2\varepsilon_{xy} \\
\varepsilon_{xx}-\varepsilon_{yy}+i2\varepsilon_{xy} & 0
\end{array}\right],
\end{equation}
\end{subequations}
where the first term $H^{1}_{0}(\bm q)$ is the first order approximation, \emph{i.e.}, the gapped Dirac Hamiltonian \cite{kp1},  the second term is the trigonal warping \cite{kpTW}, and the strain perturbation contains the third and fourth terms, \emph{i.e.}, the strain-induced scalar potential ($V_c$ and $V_v$) and pseudomagnetic vector potential ($\bm A$), with their corresponding forms as follows:
\begin{subequations}
\begin{eqnarray}
 V_{c(v)}=\alpha_{c(v)}\beta(\varepsilon_{xx}+\varepsilon_{yy}),
\end{eqnarray}
\begin{equation}
 A_x=\frac{\hbar\alpha\beta}{aet}(\varepsilon_{xx}-\varepsilon_{yy}),  A_y=\frac{\hbar\alpha\beta}{aet}(-2\varepsilon_{xy}).
\end{equation}
\end{subequations}
The obtained scalar and vector potentials in the linear elastic approximation have the similar forms as those in other 2D hexagonal crystals, such as $BN$ and graphene \cite{PFBNGrap1,PFBNGrap2,PFBNGrap3,PFBNGrap4,PFBNGrap5,PFBNGrap6,PFBNGrap7}. The previous results for the linear elastic approximation in Refs. [31-34] of the main text are also consistent with the present strain-dependent $\bm{k\cdot p}$ Hamiltonian. By fitting the energy band structures between the $\bm{k\cdot p}$ and tight-binding (TB) models for unstrained TMDMs, we obtain the energy parameters ($\Delta$, $t$, $\gamma_1$, $\gamma_2$ and $\gamma_3$) for all TMDMs involving the second order approximations, as shown in Table \ref{Stab2}. The other energy parameters ($\alpha_c$, $\alpha_v$ and $\alpha$) in the strain perturbation terms in principle can also be obtained through the energy band comparison between the DFT calculations and TB calculations for strained TMDMs. In addition, the pseudomagnetic vector potential $\bm A$ is direction-dependent. If the $x$ direction has an angle ¦È with respect to the armchair direction of TMDMs, $\bm A$ has the following general forms \cite{PFGene1,PFGene2,PFGene3,PFGene4,PFGene5,PFGene6,PFGene7}:
\begin{subequations}
\begin{eqnarray}
 A_x=\frac{\hbar\alpha\beta}{aet}[(\varepsilon_{xx}-\varepsilon_{yy})\cos(3\theta) -2\varepsilon_{xy}\sin (3\theta)],
\end{eqnarray}
\begin{equation}
 A_y=-\frac{\hbar\alpha\beta}{aet}[(\varepsilon_{xx}-\varepsilon_{yy})\sin(3\theta) +2\varepsilon_{xy}\cos(3\theta)].
\end{equation}
\end{subequations}
\begin{table}[b]
\caption{\label{Stab2}
The fitting values for energy parameters $\Delta$, $t$, $\gamma_1$, $\gamma_2$ and $\gamma_3$ (in units of eV) between the continuum approximation involving the trigonal warping and TB model for TMDMs.}
\begin{ruledtabular}
\begin{tabular}{ccccccc}
 & MoS$_2$ & MoSe$_2$ & MoTe$_2$ & WS$_2$ & WSe$_2$ & WTe$_2$ \\
\hline
$\Delta$ & 1.6579 & 1.4293 & 1.2302 & 1.8062 & 1.5412 & 1.0668 \\
$t$ & 1.0301 & 0.8752 & 0.7186 & 1.2910 & 1.0812 & 0.9462 \\
$\gamma_1$ & 0.1413 & 0.0803 & 0.1349 & 0.1783 & 0.1802 & 0.1903 \\
$\gamma_2$ & 0.0115 & 0.0203 & 0.0282 & 0.0531 & 0.0086 & 0.0857 \\
$\gamma_3$ & -0.1047 & -0.0826 & -0.0138 & -0.0853 & -0.0760 & -0.0692 \\
\end{tabular}
\end{ruledtabular}
\end{table}
\section{\label{sec3}Piezoelectric Berry curvatures in the first order continuum approximation}
In order to evaluate the piezoelectric Berry curvatures in the first order approximation (neglecting the trigonal warping), we need to first acquire the energy eigenvalues and normalized eigenstates for the conduction $(+)$ and valence $(-)$ bands of TMDMs without the strain perturbation, as follows:
\begin{subequations}
\begin{equation}
E^0_\pm=\pm\sqrt{a^2t^2(q^2_x+q^2_y)+(\Delta/2)^2},
\end{equation}
\begin{equation}
|u^0_\pm\rangle=\frac{1}{\sqrt{2E^0_+}}
\left[
\begin{array}{c}
\frac{at(q_x-iq_y)}{\sqrt{E^0_+\mp(\Delta/2)}}\\
\pm\sqrt{E^0_+\mp(\Delta/2)}
\end{array}\right].
\end{equation}
\end{subequations}
Then the partial derivatives of Hamiltonian (neglecting the trigonal warping) in Eqs. (\ref{ContHami}) with respect to $q_y$ and $\varepsilon_{yy}$ read
\begin{subequations}
\label{partialHami}
\begin{equation}
\frac{\partial H_0(\bm q)}{q_y}=\frac{\partial H^1_0(\bm q)}{q_y}=
\left(
\begin{array}{cc}
0 & -iat\\
iat & 0
\end{array}\right),
\end{equation}
\begin{equation}
\frac{\partial H_K(\bm q,\bm \varepsilon)}{\varepsilon_{yy}}=\frac{\partial H^s_S(\bm \varepsilon)}{\varepsilon_{yy}}+\frac{\partial H^v_S(\bm \varepsilon)}{\varepsilon_{yy}},
\end{equation}
\begin{equation}
\frac{\partial H^s_S(\bm \varepsilon)}{\varepsilon_{yy}}=
\left(
\begin{array}{cc}
\alpha_c\beta & 0\\
0 & \alpha_v\beta
\end{array}\right),
\end{equation}
\begin{equation}
\frac{\partial H^v_S(\bm \varepsilon)}{\varepsilon_{yy}}=
\left(
\begin{array}{cc}
0 & -\alpha\beta\\
-\alpha\beta & 0
\end{array}\right).
\end{equation}
\end{subequations}
Inserting Eqs. (\ref{partialHami}) into the Berry curvature expression in Eq. (4b) of the main text, we obtain the piezoelectric Berry curvatures $\Omega^s_{2,22}$ and $\Omega^v_{2,22}$ contributed by the corresponding scalar and vector potentials, as shown in Eqs. (5) of the main text.
\section{\label{sec4}Trigonal warping effects on the piezoelectricity of TMDMs}
The trigonal warping changes the distribution of the Berry curvatures in the first order continuum approximation near the $K$ valley and hence also contributes to the piezoelectricity of TMDMs. For the second order approximation, the energy eigenvalues and normalized eigenstates in Eqs. (\ref{ContHami}) for the conduction $(+)$ and valence $(-)$ bands read
\begin{subequations}
\begin{equation}
E^0_\pm=\frac{\Delta_++\Delta_-\pm\sqrt{4|f(q_x,q_y)|^2+(\Delta_+-\Delta_-)^2}}{2},
\end{equation}
\begin{equation}
|u^0_\pm\rangle=\frac{1}{\sqrt{2E^0_+-\Delta_+-\Delta_-}}
\left[
\begin{array}{c}
\frac{f(q_x,q_y)}{\sqrt{E^0_+-\Delta_\pm}}\\
\pm\sqrt{E^0_+-\Delta_\pm}
\end{array}\right],
\end{equation}
\end{subequations}
where $\Delta_+$, $\Delta_-$ and $f(q_x,q_y)$ have been set as
\begin{subequations}
\begin{equation}
\Delta_+=\gamma_1a^2(q^2_x+q^2_y)+(\Delta/2),
\end{equation}
\begin{equation}
\Delta_-=\gamma_2a^2(q^2_x+q^2_y)-(\Delta/2),
\end{equation}
\begin{equation}
f(q_x,q_y)=at(q_x-iq_y)+\gamma_3a^2(q_x+iq_y)^2.
\end{equation}
\end{subequations}
The partial derivative of Hamiltonian including the trigonal warping in Eqs. (\ref{ContHami}) with respect to $q_y$ reads
\begin{equation}
\frac{\partial H_0(\bm q)}{q_y}=
\left(
\begin{array}{cc}
2\gamma_1a^2q_y & \partial{f(q_x,q_y)}/\partial{q_y}\\
\partial{f^*(q_x,q_y)}/\partial{q_y} & 2\gamma_2a^2q_y
\end{array}\right),
\end{equation}
where $\partial{f(q_x,q_y)}/\partial{q_y}=-iat+i2\gamma_3a^2(q_x+iq_y)$, and $\partial{f^*(q_x,q_y)}/\partial{q_y}=iat-i2\gamma_3a^2(q_x-iq_y)$. In this case, the corresponding Berry curvatures read
\begin{subequations}
\label{TWBerrCuv}
\begin{equation}
\Omega_{2,22}=\Omega^s_{2,22}+\Omega^v_{2,22},
\end{equation}
\begin{equation}
\Omega^s_{2,22}=\frac{2a\beta(\alpha_c-\alpha_v)(t-2aq_x\gamma_3) (aq_xt+a^2q^2\gamma_3)}{[4|f(q_x,q_y)|^2+(\Delta_+-\Delta_-)^2]^{{3/2}}},
\end{equation}
\begin{equation}
\Omega^v_{2,22}=\frac{2a\alpha\beta(t-2aq_x\gamma_3)[2a^2q^2_y(\gamma_2-\gamma_1)
+(\Delta_+-\Delta_-)]}{[4|f(q_x,q_y)|^2+(\Delta_+-\Delta_-)^2]^{{3/2}}}.
\end{equation}
\end{subequations}
Then the piezoelectric coefficient $e_{222}$ reads
\begin{eqnarray}
e_{222} &=& \frac{4e}{(2\pi)^2} \int_0^{2\pi} d\theta \int_0^{q_m}(\Omega^s_{2,22}+\Omega^v_{2,22})q dq,
\end{eqnarray}
where $q = \sqrt{q_x^2+q_y^2}$, and $\pi q^2_m = S_{BZ}/2$ with $S_{BZ} = 8\pi^2/{\sqrt{3}a^2}$ in order to conserve the total number of states. Different from the first order approximation, the strain-induced scalar potential  in the second approximation also benefits the piezoelectric coefficient, because the previous odd distribution of the Berry curvature $\Omega^s_{2,22}$ in Eq. (5a) of the main text with respect to $q_x$ has been changed by the trigonal warping, as demonstrated in Eqs. (\ref{TWBerrCuv}). Consequently, the piezoelectric coefficient contains two parts contributed by both scalar and vector potentials.